# Estimation of background carrier concentration in fully depleted GaN films

Hareesh Chandrasekar,[1,] Manikant Singh,[1] Srinivasan Raghavan,[1] and Navakanta Bhat[1] [a)]

[1]*Centre for Nano Science and Engineering, Indian Institute of Science,*

*Bangalore, 560012, India*

**ABSTRACT**

Buffer leakage is an important parasitic loss mechanism in AlGaN/GaN HEMTs and hence various methods are employed to grow semi-insulating buffer layers. Quantification of carrier concentration in such buffers using conventional capacitance based profiling techniques is challenging due to their fully depleted nature even at zero bias voltages. We provide a simple and effective model to extract carrier concentrations in fully depleted GaN films using capacitance-voltage (C-V) measurements. Extensive mercury probe C-V profiling has been performed on GaN films of differing thicknesses and doping levels in order to validate this model. Carrier concentrations as extracted from both the conventional C-V technique for partially depleted films having the same doping concentration, and Hall measurements show excellent agreement with those predicted by the proposed model thus establishing the utility of this technique. This model can be readily extended to estimate background carrier concentrations from the depletion region capacitances of HEMT structures and fully depleted films of any class of semiconductor materials.

---

[a)] Author to whom correspondence should be addressed. Electronic mail:

navakant@ece.iisc.ernet.in.

INTRODUCTION

Significant advances in material quality and device designs have enabled the use of AlGaN/GaN high electron mobility transistors (HEMTs) in high-power and high-frequency applications. Parasitic losses within the device stack itself, primarily the existence of conduction pathways through the buffer layer and hetero-interfaces, play a major role in degrading device performance.[1]

Doping of GaN buffers using iron[2] and carbon[3] is commonly employed to obtain high-resistivity layers which suppress parasitic buffer leakage. The semi-insulating nature of these buffers renders them fully depleted of charge carriers. The trend to reduce the thickness of the entire device stack to form ultra-thin devices[4] also leads to the formation of fully depleted layers, similar to the fully-depleted silicon on insulator (FD-SOI) scenario. For example, record defect reduction in GaN films on Si has recently been demonstrated with a buffer layer stack as thin as 210 nm which obviates the need to grow thicker GaN.[5] These developments, in addition to their obvious advantages in reducing epitaxial cost and complexity by employing thin GaN buffers, give rise to a scenario where carrier concentration estimation is again a challenge. Mercury probe based capacitance-voltage measurements have long been used a quick, reliable and non-destructive test of the electrical properties of both the epitaxial layers individually and the complete HEMT stack[6-8]. While capacitance-voltage profiling of semiconducting films biased in the depletion region offers a simple way to extract carrier concentrations, the fully depleted nature of semi-insulating buffers gives rise to insignificant changes in capacitance with voltage. As a result, the conventional method to extract carrier density based on the change in capacitance with applied voltage cannot be employed in such cases. Typically, leakage current measurements between two contact pads placed on the buffer layer are used in order to estimate their resistivity and electrical breakdown limits.[1] This technique does not yield a quantitative estimate of the carrier concentration and involve further fabrication processes to realize. While TLM[9] and even micro-Raman[10] measurements have been employed to estimate the resistivity of these buffers and indirectly infer the carrier concentrations, the need for further fabrication steps and instrumentation are necessary to implement these methods. Therefore facile and high throughput methods to extract doping concentration from simple C-V measurements even in case of fully depleted layers are desirable.

**RESULTS AND DISCUSSION**

For a standard Schottky junction biased in the depletion region, the carrier concentration as extracted from changes in the capacitance, due to depletion width modulation with applied reverse bias, can be written as

$$N(W_D) = \frac{2}{q\varepsilon_0\varepsilon_r}\left[-\frac{1}{d(1/C_D^2)/dV}\right] \quad (1)$$

or

$$N(W_D) = \frac{1}{q\varepsilon_0\varepsilon_r}\left[\frac{C_D^3}{d(C_D)/dV}\right] \quad (2)$$

where N stands for the carrier concentration given by the net ionized shallow impurities, $W_D$ the depletion width, $C_D$ the depletion layer capacitance, V the applied voltage, q the electron charge, $\varepsilon_0$ represents the permittivity of free space and $\varepsilon_r$ the dielectric constant of the semiconductor material. The depletion width is further given by

$$W_D = \sqrt{\frac{2\varepsilon_0\varepsilon_r}{qN}(\psi_{bi} - V)} \quad (3)$$

with $\Psi_{bi}$ being the built-in voltage of the Schottky junction.

While the built-in voltage should ideally depend on the work function difference between the metal and semiconductor, presence of surface states typically leads to Fermi level pinning which could significantly alter the value of $\Psi_{bi}$. In silicon, the Fermi level is known to be pinned to a level $E_g/3$ above the valence band.[11] Fig. 1 shows the C-V curve from an n-type silicon standard wafer obtained using an MDC mercury probe in the dot-ring configuration. A uniform carrier concentration of $7.66 \times 10^{14}$ cm$^{-3}$ was extracted using Eq. (1)-(3), consistent with the known resistivity of the standard, while the built-in voltage was found to be 0.38 V as shown from the plot of $1/C^2$ against voltage in the inset. The built-in voltage of 0.38 V clearly indicates that the Fermi level is pinned in the lower third of the bandgap, as the ideal built-in voltage, from Schottky junction theory, for a work function difference between mercury (4.53 eV) and n-type silicon of $7.66 \times 10^{14}$ cm$^{-3}$ doping would be 0.22 V.

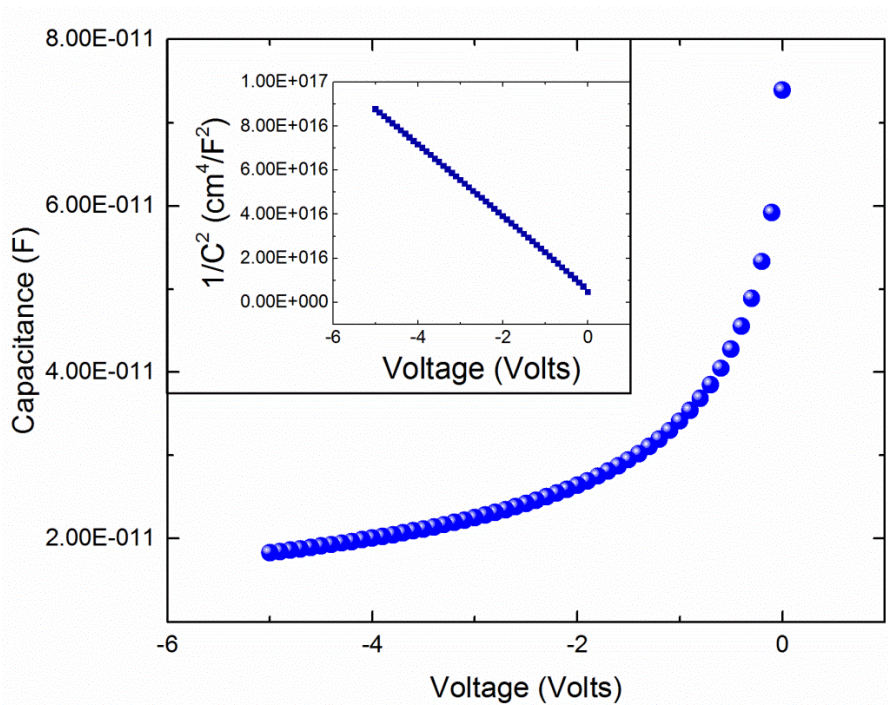

FIG. 1. Hg probe capacitance-voltage curve for a standard silicon uniformly doped wafer of doping density $7.66 \times 10^{14}$ cm$^{-3}$. The inset shows the corresponding $1/C^2$ vs V plot with the x-intercept giving the built-in voltage of 0.37 V.

The GaN samples used in this study were grown by MOCVD on silicon substrates using an AlN nucleation layer,[12] stress mitigating transition layers of step graded AlGaN layers (75%, 50% and 25% Al composition)[13] and an AlN interlayer prior to GaN growth, as shown in Fig. 2 (a). n-type doping of the GaN buffers were performed using silane as the Si precursor. C-V measurements were performed using a mercury probe connected to an Agilent 4294A impedance analyzer at a frequency of 5 kHz to reduce series resistance and dispersion effects.[8] The dissipation factor is a measure of the series resistances present in the circuit during a capacitance measurement and is defined as $D=\omega CR_S$ for the series resistance-capacitance model. Lower values of D are indicative of more accurate measured capacitances. All D values during the capacitance measurements in this study were less than 0.1 thus ensuring that the measured characteristics can be reliably analyzed.[14] The C-V for a 500 nm undoped GaN film thus grown (Fig. 2(b)) shows no change in capacitance with applied reverse voltage clearly indicating that the film is fully depleted even at 0 V bias which renders any attempt to extract the doping concentration using Eq. (1)-(3) meaningless. In the buffers layers measured in this study, the variation in capacitance between low and high measurement frequencies (5 kHz and 1 MHz) change by less than 5% as also shown in the Fig. 2(b) below for the undoped 500 nm sample. This is unsurprising since the buffer layers themselves are highly resistive and hence any series resistance effects due to the mercury probe contact do not play a major role.[8] If, on the other hand, a HEMT stack is now fabricated on these buffers, series resistance and capacitance dispersion effects due to the series resistance will indeed play a role and needs to be corrected as detailed by Shealy and Brown[8].

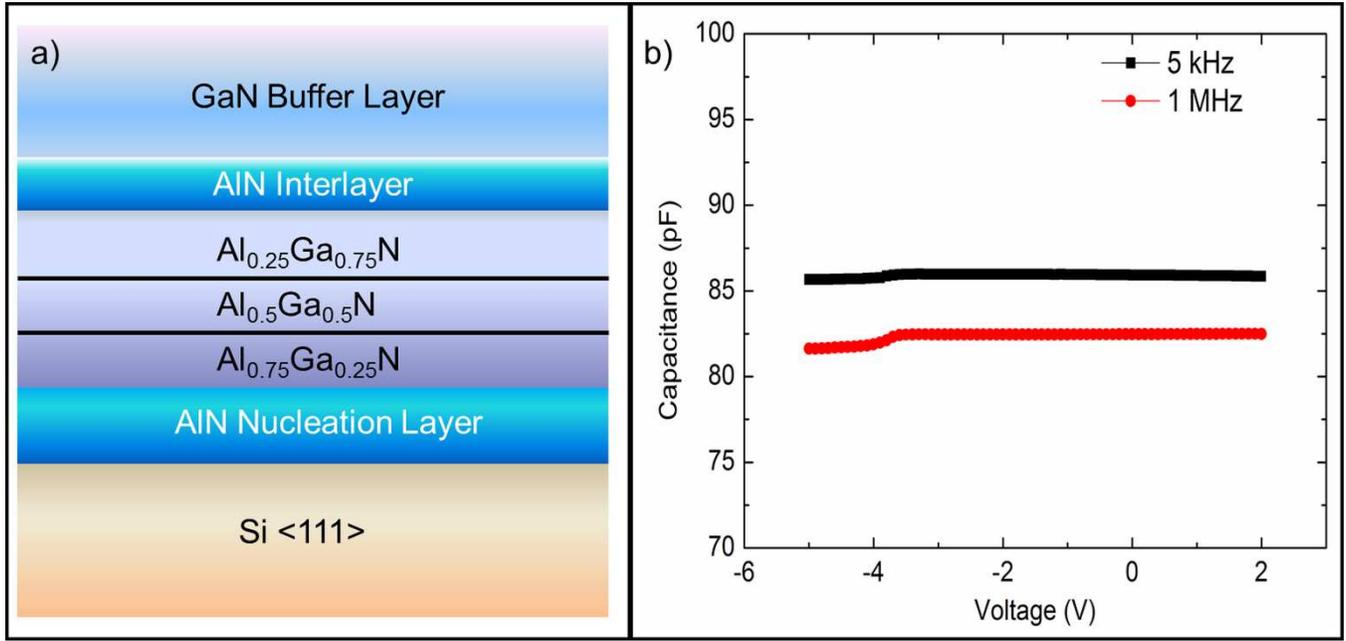

FIG. 2. a) Layer stack of the GaN buffers on silicon used in this study. b) Hg probe C-V characteristics of the 500 nm undoped GaN buffer, at 5 kHz and 1 MHz, illustrating the fully-depleted nature of the film.

In order to estimate the doping density in this case, we note that the ratio of capacitances at 0 V between the Si standard and the 500 nm GaN film is 1.142. Further, we postulate that this ratio is essentially related to the material properties of both semiconductors from the Schottky junction theory.

$$C_j = \sqrt{\frac{q\varepsilon_0 \varepsilon_S N_d}{2(\psi_{bi} - V_{app})}} \qquad (4)$$

where is the junction capacitance of the Schottky junction formed, $N_d$ is the net donor density, $\psi_{bi}$ is the built-in voltage and $V_{app}$ is the applied voltage. We see that the ratio of capacitances of GaN and Si now becomes,

$$\frac{C_{GaN}}{C_{Si}} = \sqrt{\frac{\varepsilon_{GaN} N_{GaN}}{\psi_{bi,GaN}} \times \frac{\psi_{bi,Si}}{\varepsilon_{Si} N_{Si}}} \qquad (5)$$

Substituting all the relevant parameters, we obtain

$$\frac{N_{GaN}}{\psi_{bi,GaN}} = 2.285 \times 10^{15} \left(\frac{C_{GaN}}{C_{Si}}\right)^2 cm^{-3}V^{-1} \qquad (6)$$

with $\varepsilon_{GaN}$ = 10.6[15], $\varepsilon_{Si}$ = 11.7 and $n_{Si}$ and $\psi_{Si}$ are $7.66 \times 10^{14}$ cm$^{-3}$ and 0.37 V as calculated earlier. Since the film is fully depleted for a thickness of 500 nm, the depletion width in this case is at least equal to the film thickness, $t_f$ i.e., $W_D^2 \geq t_f^2$ which gives

$$\frac{2\varepsilon_0 \varepsilon_{GaN} \psi_{bi,GaN}}{qN_{GaN}} \geq t_f^2 \qquad (7)$$

$$\frac{N_{GaN}}{\psi_{bi,GaN}} \leq \frac{2\varepsilon_0 \varepsilon_{GaN}}{qt_f^2} \; cm^{-3}V^{-1} \qquad (8)$$

This gives an upper bound for the doping concentration. We can now solve for $n_{GaN}$ subject to the constraints imposed by (5) and (8). Assuming that the Fermi level is pinned 1.0 eV below the conduction band, the model proposed in Eq. (5) or (6) gives us a carrier concentration of $3 \times 10^{15}$ cm$^{-3}$ and Eq.(8) sets an upper bound of $4.69 \times 10^{15}$ cm$^{-3}$ for the 500 nm GaN film. It is important to note that this analysis in its present form is valid only when the applied voltage, and hence the depletion region, is confined to the GaN layer alone so that the transition layers below do not also contribute to the measured capacitance. Introducing an AlN interlayer acts as an effective electrical barrier between the GaN film and the transition region for the range of voltages used in this study. In order to verify that the AlN interlayer does indeed confine the depletion widths and hence measured capacitance to the GaN buffer layer, we model the electrostatics of this stack including the two mercury Schottky contacts using the Silvaco ATLAS device simulator. The simulated capacitance-voltage curves of a 500 nm and 1000 nm thick GaN film with a nominal dopant density of $5 \times 10^{15}$ cm$^{-3}$ donors are as shown in Fig. 3. We see that the capacitance halves when the GaN film thickness is doubled from 500 nm to 1000 nm. This clearly indicates that the contribution to the measured capacitance is from the GaN film alone and the effect of the AlN interlayer is analogous to the current blocking layer used in LEDs for example. We see that despite the use of the AlN interlayer in this study, the model proposed here can be extended in a straightforward manner to any combination of transition layers just by simple simulations to determine the extent of depletion region penetration, based on band offsets and carrier concentrations of the layers used. For example, while we have assumed a single uniformly doped GaN layer here, the case of GaN stacks with different doping concentration can be considered as capacitances in series. Also, in case of non-uniform doping profiles, the capacitance changes from $C=C_{J0}(V-\psi_{bi})^{-1/2}$ to a $C_{J0}(V-\psi_{bi})^{-m}$ dependence where the value of m depends on the exact doping profile. For example m is 1/3 for linearly graded doping profiles from simple p-n junction theory.[16] Also, when the type of doping is varied from p to n, the p-n junction capacitance is hence expected to serve the same purpose as the AlN interlayer we use in the present study as a "potential step". So effectively a p-n junction would actually confine all capacitance measured to the layers above it which can then be handled as described in the preceding statements.

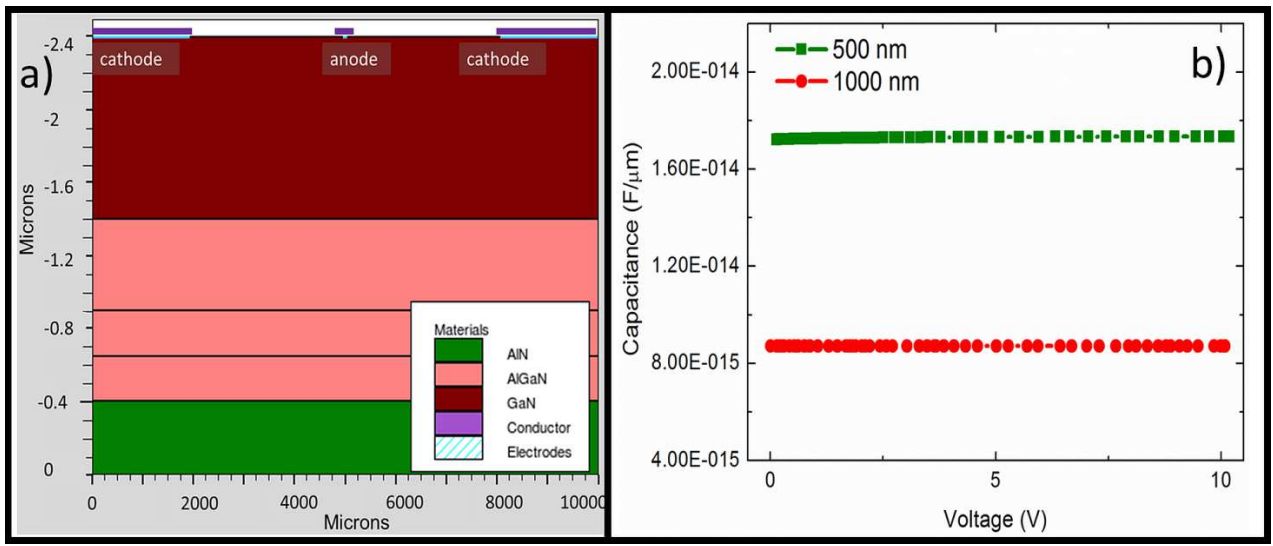

FIG. 3. a) Schematic of the layer stack used in the simulation, b) Capacitance per unit width (F/μm) with voltage indicating the fully depleted nature of the films (evident from the <1% change in y-axis scale) for undoped GaN buffers with nominal charge density of $5\times10^{15}$ cm$^{-3}$ and thicknesses of 500 nm and 1000 nm. We see that the capacitance halves for the 1000 nm case ($8.71\times10^{-15}$ F/ μm) as compared to the 500 nm film ($1.71\times10^{-14}$ F/ μm) indicating that the depletion width is restricted to the GaN buffer as discussed in the text.

At this juncture it is pertinent to point out that the proposed method estimates the *mobile* background carrier concentration in the GaN layers and *not* the background impurity density. It has previously been reported by Uren et al.[17] that despite the low mobile carrier density obtained from C-V, net deep acceptor densities of $\sim 10^{16}$ cm$^{-3}$ are present in the insulating GaN buffers, and indeed are essential to eliminate punch-through effects in short-channel HEMTs. If we assume a net GaN carrier density of $10^{16}$ cm$^{-3}$ due to background doping in the undoped GaN layers, we see that the Fermi level of the GaN buffer, which can be readily estimated using elementary semiconductor statistics as $E_C - E_F = kT\ln(N_C/n) = 0.141$ eV, with an $N_c$ of $2.3 \times 10^{18}$ cm$^{-3}$ for wurtzite GaN. Following Uren et al.,[17] we assume that deep acceptors are present $\sim 0.9$ eV below the conduction band. This results in a large difference in $E_F - E_T$ of 0.76 eV which would mean that almost all the trap states are occupied by electrons and as such these electrons do not contribute to any conduction unless they are de-trapped. Since the de-trapping mechanism is thermally activated, and $E_F - E_T$ is extremely large compared to kT, electrons trapped in such deep acceptor states are not expected to significantly contribute to conduction. While this underestimates the *impurity* concentration in the buffer, we wish to emphasize again that the proposed model quantifies the mobile background carriers and not the background impurity density which is also composed of deep trap states.

In order to elucidate our proposed technique further, we present C-V measurements performed on a range of film thicknesses and doping concentrations. Fig. 4 shows the measured C-V curves for three different doping concentrations of GaN for different film thicknesses in each case. The unintentionally doped sample (u-GaN) does not have any dopant gas flow during growth but is intrinsically n-doped. On the other hand, moderately doped sample n1 and n2 are intentionally doped with silane flow rates of 0.178 nmol/min and 0.884 nmol/min respectively during growth.

As discussed earlier, fully depleted films can arise due to two considerations – low values of mobile carrier concentrations and low values of thickness, as both these reasons lead to a scenario where film thickness < depletion width. The transition from fully depleted to the partially depleted regime with an increase in thickness and doping concentration is clearly visible as expected. The undoped GaN samples remain fully depleted for even film thicknesses of up to 1μm as shown in Fig. 4(a). We also note that the measured capacitance of the 500 nm film is exactly twice that for the 1000 nm film thus confirming the simulation results about the depletion width being restricted to the topmost GaN buffer. Both the Si-doped GaN samples, n1 and n2, clearly show fully depleted C-V curves for 100 nm films which transition to the partially depleted scenario upon raising the film thickness to 500 nm and 750 nm seen in Figs. 4 (b) and (c). Since we now have both fully depleted and partially depleted films for a given doping concentration, just by varying the thickness, the doping density as extracted from the proposed model for the fully depleted films can now be compared to that from the conventional C-V analysis.

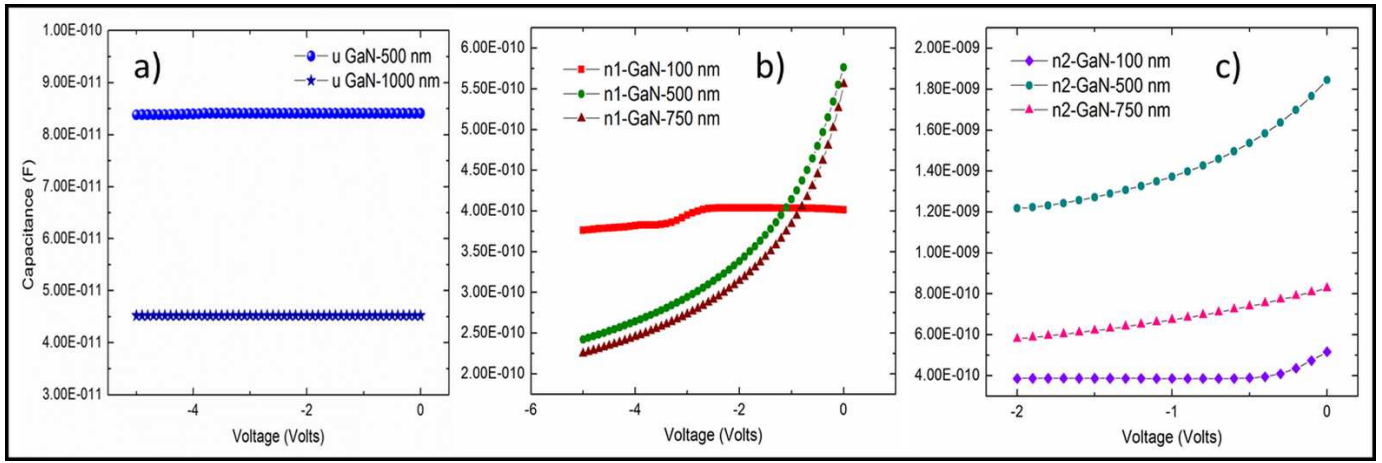

FIG. 4. Hg probe capacitance-voltage curves for different GaN buffer doping-thickness combinations indicating the transition from fully depleted to partially depleted films with increasing doping and thicknesses for a) unintentionally doped GaN of 500 and 1000 nm thickness, b) lightly doped GaN of 100 nm, 500 nm and 750 nm thickness and c) moderately doped GaN of 100 nm, 500 nm and 750 nm thicknesses.

Furthermore, Hall bar structures were fabricated and tested on all the doped GaN films across an entire 2" wafer in order to obtain an independent measure of the carrier concentration that can then be compared to those extracted from both the C-V techniques. Fig. 5 shows the carrier concentrations as extracted from C-V using the proposed (assuming a built-in voltage of 1V) and conventional techniques

(where possible) and as measured using the Hall bars. The upper bound on the carrier density set by the depletion width consideration of Eq. (8) is also indicated for the fully depleted films. It should be noted that thickness variations of ±5% across the wafer are observed in growth which in turn leads to a 10% error in estimation of the upper bounds indicated here.

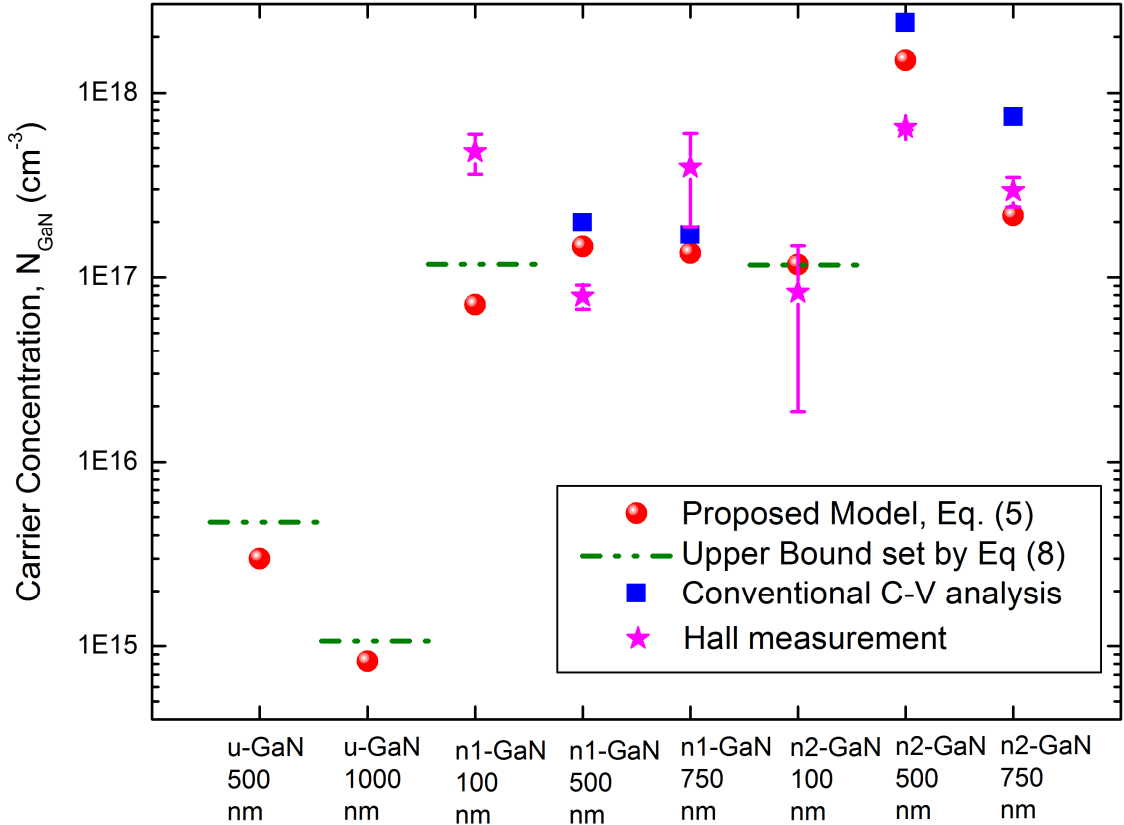

FIG. 5. Carrier concentrations of the undoped and Si-doped GaN buffers used in this study as extracted from the proposed model (circles), conventional C-V analysis for the partially depleted films (squares) and from Hall measurements (stars). The error bars are indicative of standard deviation from at least 5 Hall bar structures measured across an entire 2" wafer. The upper bound for the fully depleted films from the depletion width consideration of Eq. (8) is also indicated. We see that the proposed technique shows very good agreement with those measured from the other two methods for partially depleted structures and is applicable for even fully depleted films where these techniques fail.

The following key points from Fig. 5 are noteworthy:

1. For the fully-depleted undoped samples (u-GaN) of 500 nm and 1000 nm, the proposed model gives carrier concentration values of $3 \times 10^{15}$ cm$^{-3}$ and $8.3 \times 10^{14}$ cm$^{-3}$ respectively. We see that these values are within the upper bound set by the depletion width. Since the film is fully depleted even for a thickness of 1000 nm, the true carrier concentration in the 500 nm film is also expected to be lower than $8.3 \times 10^{14}$ cm$^{-3}$. Hence applying the proposed model overestimates the carrier density, if at all. This low background density also makes Hall measurements challenging due to the difficulty of achieving ohmic contacts to these samples.

2. The 500 nm and 750 nm films for both the Si doping levels, n1 and n2, are partially depleted and hence their C-V characteristics can also be analyzed using the conventional C-V techniques of Eq.(1)-(2). We see that the carrier concentration values extracted using the proposed model, the conventional C-V method and Hall measurements match closely. The built-in voltages for these films as extracted from the $1/C^2$-V curves are 1.03, 1.05, 1.25 and 1.8 V for the 500 and 750 nm films of n1 and n2 doping levels respectively. This supports our assumption of the built in voltages being ~1V in the fully-depleted samples. We see that the close match between the measured Hall carrier concentrations, which are DC measurements, and those estimated by the C-V model, which are high-frequency measurements, also reinforces the fact that the deep traps do not contribute significantly towards conduction.

3. 100 nm films for both the Si doping concentrations are fully depleted and the values extracted using Eq. (5) are $7.1 \times 10^{16}$ and $1.2 \times 10^{17}$ cm$^{-3}$ respectively. These values match reasonably well with the Hall measurements. We also note that for samples of a given doping concentration, the carrier concentration values extracted from the proposed model for the fully depleted films are comparable to those in the partially depleted films and well within the experimental variations in the doping levels.

4. Finally we would like to comment on the variability of the doping levels observed above. We note that even though the silane partial pressures were kept constant for all samples of a given doping density (n1 or n2), the variations in the electron concentrations, as measured by any of the techniques shown in Fig. 5, can be attributed to the interplay between dislocations acting as electron traps[18] and their reduction with increasing thickness due to dislocation bending[19], in addition to the run-to-run variability and minor thickness variation across the entire wafer over which the Hall bars were tested.

We see that the measured built-in voltages of ~1 V are consistent with the pinning of the Fermi level in GaN. In case of an ideal Hg-GaN Schottky junction, with an electron density of $10^{16}$ cm$^{-3}$, the built-in voltage based on the difference in work functions should be 0.289 V as shown in Fig 6.

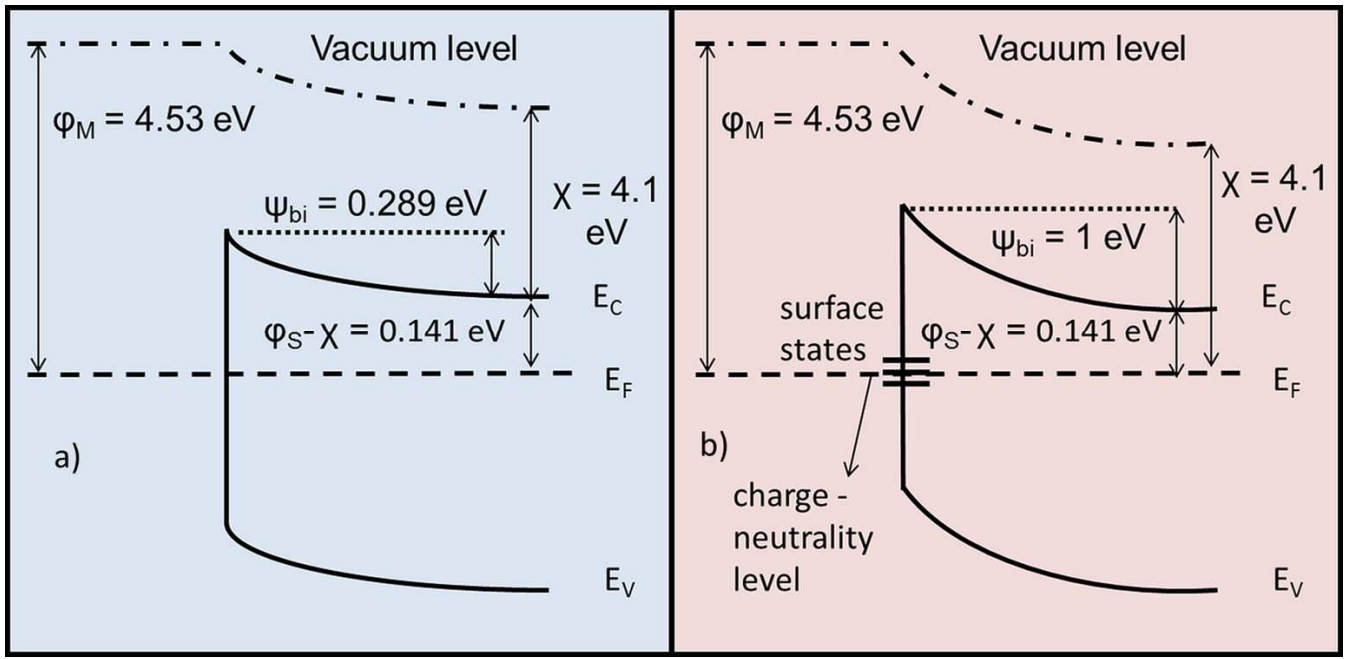

FIG. 6. Equilibrium band diagrams for the Hg-GaN junction in the case of (a) ideal Schottky junction with a built-in voltage of 0.289 V, and (b) with Fermi-level pinning due to surface states, giving rise to a built-in voltage of 1 V.

Such Fermi level pinning in GaN films has been predicted to occur, on the basis of theoretical band structure calculations, at 0.5-0.7 eV below the conduction band[20] and experimentally observed to be 1eV below Ec in case of GaN-Pt junctions.[21] While carbon in the Nitrogen site in GaN is known to form a deep acceptor state at 0.9 eV below the conduction band, dopants/deep traps in the bulk are not expected to lead to Fermi level pinning, which is essentially a surface related phenomenon.

## CONCLUSIONS

In conclusion, a method to extract the carrier concentration of fully depleted GaN films based on the Schottky junction theory has been proposed. Mercury probe capacitance measurements have been performed on GaN films of different thicknesses and doping levels to validate this model. The carrier densities extracted using the proposed technique show good agreement with those from the conventional C-V analysis for partially depleted films and Hall measurements. This simple and non-destructive technique to quantify carrier concentrations can be extended in a straightforward manner to fully depleted films of any semiconductor and to capacitances measured from the depletion region of entire HEMT stacks to evaluate the buffer layers used.

## ACKNOWLEDGEMENTS

We acknowledge the Ministry of Defence, Government of India, through sanction number TD-2008/SPL-147 for funding to carry out this research, the National Nano Fabrication Centre and the Micro and Nano Characterization Facility of the Centre for Nano Science and Engineering, Indian Institute of Science for providing access to device fabrication and electrical characterization facilities respectively. We also thank DeitY, Govt. of India for funding support through the Centre for Excellence in Nanoelectronics Phase II program.